\documentclass[pra,twocolumn, aps, preprintnumbers,superscriptaddress]{revtex4-2}
\usepackage[colorlinks=true,citecolor=blue,urlcolor=black]{hyperref}
\usepackage{verbatim}
\usepackage{amsmath}
\usepackage{latexsym}
\usepackage{revsymb}
\usepackage{yfonts}
\usepackage{ifthen}

\usepackage{natbib}
\usepackage{amsfonts}
\usepackage{amsmath}
\usepackage{amssymb}
\usepackage{amsthm}
\usepackage{graphicx}
\usepackage{bm}
\usepackage{bbm}
\usepackage{epsfig,color,amssymb}
\usepackage{subfigure}
\usepackage{amsfonts}
\usepackage{amscd}
\usepackage{amsmath}
\usepackage{multirow}
\usepackage{chemarrow}
\usepackage{dcolumn}
\usepackage{bm}
\usepackage{graphicx}
\usepackage{enumerate}
\usepackage{epsfig}
\usepackage{subfigure}
\usepackage{xcolor}
\usepackage{multirow}
\usepackage{ulem}
\usepackage{braket}
\usepackage{comment}
\usepackage{enumitem}
\usepackage{amsthm}
\renewcommand{\emph}[1]{\textit{#1}}

\begin{document}
	
	\title{High key rate quantum conference key agreement with unconditional security}
	
	\author{Xiao-Yu Cao}
	\author{Yu-shuo Lu}
	\author{Zhao Li}
	\author{Jie Gu}	
	\author{Hua-Lei Yin}\email{hlyin@nju.edu.cn}
	\author{Zeng-Bing Chen}\email{zbchen@nju.edu.cn}
	\affiliation{National Laboratory of Solid State Microstructures, School of Physics and Collaborative Innovation Center of Advanced Microstructures, Nanjing University, Nanjing 210093, China}
	
	
	\begin{abstract}
	 Quantum cryptography is a major ingredient of the future quantum internet that promises various secure communication tasks. Quantum conference key agreement (CKA) is an important cryptographic primitive of quantum cryptography, which provides the conference key among multiple users simultaneously. However, quantum CKA is currently far from practical application due to the low conference key rate. Here, we propose a quantum CKA protocol of three users with information-theoretic security. Our protocol only requires  phase-randomized weak coherent sources and threshold single-photon detectors, and is anticipated to be experimentally demonstrated over 600 km under current technology. Our scheme can be widely implemented in the approaching large-scale quantum network.
	\end{abstract}

	\maketitle
\section{INTRODUCTION}

Security of various communication tasks can be guaranteed by quantum internet, which consists of numerous quantum cryptographic networks.
Quantum key distribution (QKD) allows two remote users to share unconditionally secure key based on quantum laws~\cite{bennett1984quantum,ekert1991quantum}, which has stepped into a stage where an integrated network from satellite to ground is waiting to be built. Nevertheless, there are kinds of multiparty cryptographic primitives beyond two users, e.g., conference key agreement (CKA)~\cite{diffie1976new,burmester1994secure}. 
Quantum CKA aims to allow at least three users to share the conference key with unconditional security, which can be applied in various practical situations, such as netmeeting, online education and telemedicine. The conference key promises group encryption for legitimate users, in which any member of that group can decipher the message. As it generalizes the QKD to multiple users, it is also called multipartite QKD~\cite{matsumoto2007multiparty}.

Quantum CKA has been a research hot topic~\cite{Glaucia:2020:Quantum}. At the very beginning, it enables multiple users to share the conference key with information-theoretic security by employing multipartite entanglement~\cite{bose1998multiparticle,cabello2000multiparty,chen2007multi}. 
Besides, the measurement-device-independent CKA protocol~\cite{fu2015long} has been proposed via post-selected Greenberger-Horne-Zeilinger (GHZ) entanglement~\cite{lo1999unconditional,Shor2000Simple}. It has been generalized extensively to various cases including finite size~\cite{chen2016biased,chen2017asymmetric}, continuous variables~\cite{wu2016continuous,ottaviani2019modular} and four users with $W$ state~\cite{zhu2015w}.
Moreover, the finite-key analysis with composable security~\cite{grasselli2018finite}, device independence~\cite{ribeiro2018fully,holz2019genuine} and other special cases~\cite{li2018quantum,pivoluska2018layered,jo2019semi} are also considered in quantum CKA. The experimental demonstration of quantum CKA has been implemented recently over 50 km fiber using the state-of-the-art four-photon GHZ entanglement source~\cite{proietti2021experimental}. By directly distributing GHZ entanglement states, quantum CKA has been proved to drastically outperform QKD in resource expense for group communication~\cite{epping2017multi}.

Although progress on quantum CKA has been made, the low conference key rate, short transmission distance, and excessive resource cost seriously restrict its real life applications. It has been shown recently that the conference key rate is rigorously limited by the entanglement state distribution capacity of quantum networks~\cite{das2019universal,takeoka2019multipartite,pirandola2019general}. In order to increase the conference key rate, one may utilize quantum repeaters~\cite{duan2001long} and multi-user scheme~\cite{fu2015long} of adaptive measurement-device-independence~\cite{azuma2015all}. However, these approaches remain difficult to be implemented.

Introducing an intermediate node to perform a single-photon interference,
twin-field QKD~\cite{lucamarini2018overcoming} and its variants~\cite{Wang:2018:Twin,Lin:2018:Simple,Ma:2018:Phase,yin2019measurement,Cui:2019:Twin,Curty:2019:Simple,maeda2019repeaterless,Yin:2019:coherent,Xu:2020:Sending} are able to break the limit of the secret key capacity of quantum channels~\cite{takeoka2014fundamental,pirandola2017fundamental}, specifically, the Pirandola-Laurenza-Ottaviani-Banchi (PLOB) bound~\cite{pirandola2017fundamental}. Exploiting the twin-field theory~\cite{lucamarini2018overcoming}, some new quantum CKA protocols~\cite{grasselli2019conference,Zhao:2020:CKA,cao2021coherent} are proposed to improve the conference key rate and transmission distance. However, the scheme of Ref.~\cite{grasselli2019conference} can not be experimentally demonstrated with current technology and the key rate of Ref.~\cite{Zhao:2020:CKA} is relatively low. 
In addition, zero-error attack on coherent one-way QKD has been proposed~\cite{gonzalez2020upper}, therefore the security of Ref.~\cite{cao2021coherent} employing  the idea of coherent one-way QKD~\cite{stucki2005fast}, can not be guaranteed.

Here, we propose a three-party quantum CKA protocol inspired by the twin-field theory with vacuum state and single-photon encoding~\cite{Wang:2018:Twin,yin2019measurement}, i.e., the sending-or-not-sending scheme~\cite{Wang:2018:Twin}.
The conference key rate of our protocol scales with the square-root of the total channel transmittance, which enables our scheme to break the limit of entanglement state distribution capacity of quantum networks, and our protocol can be experimentally demonstrated over 600 km with developed technology in twin-field QKD~\cite{minder2019experimental,Wang:2019:Beating,Zhong:2019:Proof,Liu:2019:Exp,chen2020sending,fang2020implementation}. The key rate of our protocol is three orders of magnitude higher than that of Ref.~\cite{Zhao:2020:CKA} within a range of 600 km and can remain at a high level even with a large misalignment rate of quantum channels.
	
\begin{figure}
\centering
\includegraphics[width=8cm]{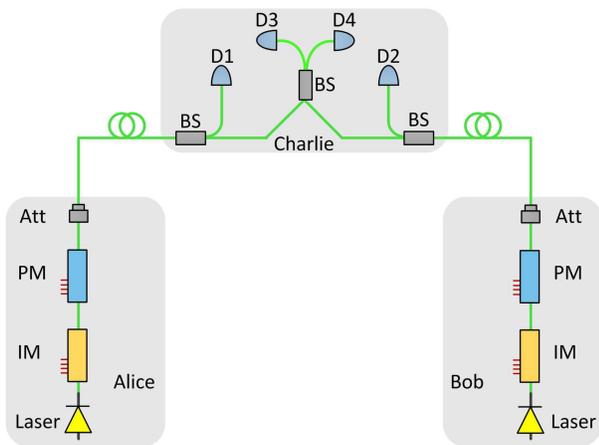}
\caption{The setup of quantum CKA protocol. Alice and Bob exploit continuous-wave lasers to generate the global phase stabilized coherent light. They employ intensity modulator (IM), phase modulator (PM) and attenuator (Att) to implement the pulse preparation, decoy-state modulation, phase randomization, phase encoding and weak-light modulation. Charlie directly measures the pulses sent by Alice and Bob with detectors D1 and D2, respectively, as the $Z$ basis measurement. Only detector D1 (D2) click represents logic bit value 0 (1), which means that a quantum state $\ket{1}_{a}\ket{0}_{b}$ ($\ket{0}_{a}\ket{1}_{b}$) has been detected. Charlie performs an interference with beam splitter (BS) and detectors D3 and D4, as the $X$ basis measurement. Only detector D3 (D4) click denotes logic bit value 0 (1), which means that an entangled state $\ket{\psi^{+}}_{ab}=\frac{1}{\sqrt{2}}(\ket{1}_{a}\ket{0}_{b}+\ket{0}_{a}\ket{1}_{b}) (\ket{\psi^{-}}_{ab}=\frac{1}{\sqrt{2}}(\ket{1}_{a}\ket{0}_{b}-\ket{0}_{a}\ket{1}_{b}))$ has been acquired. To simplify, Charlie applies a passive-basis choice.} \label{setup}
\end{figure}

\section{Protocol description}\label{sec2}
As depicted in Fig.~\ref{setup}, our scheme employs the same devices as twin-field QKD~\cite{lucamarini2018overcoming,minder2019experimental,chen2020sending,fang2020implementation}. We choose the decoy-state quantum CKA with three intensities to simplify the experiment. Hereafter, $\ket{0}$ and $\ket{1}$ form the photon number Hilbert space for vacuum and single photon. We exploit the post-selected phase-matching method to acquire the effective events with $-\delta+r\pi\leq\theta_{a}-\theta_{b}-\varphi_{ab}\leq\delta+r\pi (\rm{mod}\ 2\pi)$, where $\theta_{a}$ (for Alice) and $\theta_{b}$ (for Bob) are the random global phases and $r=0,1$. $\delta$ is a selected value and $\varphi_{ab}$ is the phase difference between the reference frames of Alice and Bob, which can be obtained by strong light reference~\cite{chen2020sending,fang2020implementation}.
The successful probability of post-selected phase-matching is $p_{\rm pm}=\frac{2\delta}{\pi}$ under the assumption of continuous phase randomization.
Our protocol has three legitimate users, Alice, Bob and Charlie. Here, we consider a symmetric case. The detailed protocol is described as follows.

{\it{1. Preparation.}}~Alice and Bob independently and randomly choose the $Z$ and $X$ bases. Alice prepares phase-randomized weak coherent pulses using intensities 0 and $\mu$, with probability $1-t$ and $t$, for logic bit $0$ and $1$ in the $Z$ basis. Bob does the same. Alice (Bob) generates phase-randomized weak coherent pulses $\ket{e^{i(\theta_{a}+x_{a}\pi)}\sqrt{k_{a}}}$ ($\ket{e^{i(\theta_{b}+x_{b}\pi)}\sqrt{k_{b}}}$) for logic bit $x_{a}$ ($x_{b}$) in the $X$ basis, with $x_{a},x_{b}\in\{0,1\}$, $\theta_{a},\theta_{b}\in[0,2\pi)$ and $k_{a},k_{b}\in\{\mu,\nu,0\}$. They send the optical pulses to Charlie through the insecure quantum channels.

{\it{2. Measurement.}}~Charlie randomly measures the received pulse pairs in the $Z$ and $X$ bases. For the $Z$ basis, one and only one click between detectors D1 and D2 reveals a successful event.
For the $X$ basis, one and only one click between detectors D3 and D4 indicates a successful outcome.

{\it{3. Reconciliation.}}~For each successful measurement event, Alice, Bob, and Charlie announce the basis information. Alice and Bob publish the intensity information unless Alice, Bob, and Charlie all select the $Z$ basis.
Alice and Bob disclose the global phase and perform post-selected phase-matching when they both choose intensity $\nu$ and Charlie selects the $X$ basis. All classical communications among Alice, Bob, and Charlie are transmitted via the authenticated classical channels.
Alice always flips her logic bit if choosing the $Z$ basis. She will flip her logic bit if selecting the $X$ basis and $r=1$.

{\it{4. Parameter estimation.}}~Alice, Bob, and Charlie adopt the data of the $Z$ basis as the raw key. The data of the $X$ basis is exploited to estimate the leaked information. They exploit the decoy-state method~\cite{wang2005beating,lo2005decoy} to the yield and phase error rate of joint single-photon state $\rho_{AB}^{1}=\frac{1}{2}(\ket{01}_{ab}\bra{01}+\ket{10}_{ab}\bra{10})$.

{\it{5. Postprocessing.}}~They extract the common conference key using classical error correction, error verification and privacy amplification.

\section{Security analysis}

For the case of the $Z$ basis, the joint quantum states $\ket{01}_{ab}$ and $\ket{10}_{ab}$ form the joint single-photon state $\rho_{AB}^{1}=\frac{1}{2}(\ket{01}_{ab}\bra{01}+\ket{10}_{ab}\bra{10})$ with probability $2t(1-t)\mu e^{-\mu}$, which can be utilized to extract key~\cite{gottesman2004security} since the coherent states prepared by Alice and Bob are phase-randomized.
The joint quantum state $\ket{01}_{ab}$ ($\ket{10}_{ab}$) can be acquired by Alice sending vacuum  (single-photon) state and Bob sending single-photon (vacuum) state.

For the case of $X$ basis,
Alice and Bob implement post-selected phase-matching and only consider the case of single photon component~\cite{yin2019measurement}, such as
\begin{equation}
\begin{aligned}\label{eq2}
\ket{e^{i\theta_{a}}\sqrt{\nu}}_{a}\ket{\pm e^{i\theta_{b}}\sqrt{\nu}}_{b}\xrightarrow{\textrm{one photon}}\frac{\ket{10}_{ab}\pm e^{i(\theta_{b}-\theta_{a})}\ket{01}_{ab}}{\sqrt{2}},
\end{aligned}
\end{equation}
where $\theta_{a},\theta_{b}\in[0,2\pi)$. The joint quantum states $\ket{\psi^{\pm}(\theta)}_{ab}=(\ket{10}_{ab}\pm e^{i\theta}\ket{01}_{ab})/\sqrt{2}$ can also form the joint single-photon state $\rho_{AB}^{1}=\frac{1}{2}(\ket{\psi^{+}(\theta)}_{ab}\bra{\psi^{+}(\theta)}+\ket{\psi^{-}(\theta)}_{ab}\bra{\psi^{-}(\theta)})$.

Here, we will exploit the entanglement distillation argument of the GHZ state~\cite{lo1999unconditional,Shor2000Simple,fu2015long} to provide security against coherent attacks.
We introduce a virtual protocol to prove the security of practical protocol as shown in Fig.~\ref{setup}. 
For the joint single-photon state case, no one can distinguish the practical protocol and the virtual protocol P1, which is similar to the twin-field QKD~\cite{Wang:2018:Twin,yin2019measurement}.

\emph{Virtual protocol P1}. Let P1 represent a virtual entanglement-based protocol. 
(i) Alice prepares an entangled state
\begin{equation}
	\begin{aligned}
		\ket{\Psi}_{Aa}=&\sqrt{1-t}\ket{0}_{A}\ket{0}_{a}+\sqrt{t}\ket{1}_{A}\ket{1}_{a}\\
		=&\frac{1}{\sqrt{2}}[\ket{+}_{A}(\sqrt{1-t}\ket{0}_{a}+\sqrt{t}\ket{1}_{a})\\
		&+\ket{-}_{A}(\sqrt{1-t}\ket{0}_{a}-\sqrt{t}\ket{1}_{a})]
	\end{aligned}
\end{equation}
where $A$ and $a$ are the qubit system and optical mode, respectively. Note that $\ket{\pm}_{A}=(\ket{0}_{A}+\ket{1}_{A})/\sqrt{2}$. Likewise, Bob does the same. A sifting step is needed to be performed, i.e.,  Alice and Bob only reserve the case that the total photon number is one between optical modes $a$ and $b$. After this step, the normalized joint state is
\begin{equation}
	\begin{aligned}
\ket{\phi}_{ABab}=&\frac{1}{\sqrt{2}}(\ket{10}_{AB}\ket{10}_{ab}+\ket{01}_{AB}\ket{01}_{ab})\\
=&\frac{1}{\sqrt{2}}\frac{\ket{++}_{AB}-\ket{--}_{AB}}{\sqrt{2}}\frac{\ket{10}_{ab}+\ket{01}_{ab}}{\sqrt{2}}\\
&+\frac{1}{\sqrt{2}}\frac{\ket{+-}_{AB}-\ket{-+}_{AB}}{\sqrt{2}}\frac{\ket{10}_{ab}-\ket{01}_{ab}}{\sqrt{2}}.
	\end{aligned}
\end{equation}
Then, Alice (Bob) sends optical mode $a$ ($b$) to Charlie and keeps the qubit system $A$ ($B$). Let optical modes $\ket{10}_{ab}$ and $\ket{01}_{ab}$ denote the qubit system $\ket{0}_{C}$ and $\ket{1}_{C}$. Therefore, the joint quantum state among Alice, Bob and Charlie can be written as a GHZ state, $\ket{\phi}_{ABC}=\frac{1}{\sqrt{2}}(\ket{100}_{ABC}+\ket{011}_{ABC})$. 
(ii) Alice, Bob and Charlie independently and randomly choose the $Z$ basis or the $X$ basis to measure the received optical modes or kept qubit system. 
(iii) They disclose the basis information. 
(iv) They utilize data of the $Z$ basis as raw key while data of the $X$ basis are used to estimate Eve's eavesdropping~\cite{fu2015long}. 
(v) They generate final key with classical post-processing.

\noindent
\textbf{Practical protocol.}
The asymptotic conference key rate of the practical protocol, as shown in Fig.~\ref{setup}, can be written as~\cite{Wang:2018:Twin,yin2019measurement,Xu:2020:Sending}
\begin{equation}
\begin{aligned}
R=2t(1-t)\{e^{-\mu}Y_{0}^{z}+\mu e^{-\mu}Y_{1}^{z}[1-h(e_{1}^{x})]\}-\lambda_{\rm EC},
\end{aligned}
\end{equation}
where $Y_{0}^{z}$ is the yield when both Alice and Bob choose the $Z$ basis and send vacuum state. $\lambda_{\rm EC}=Q^{z}fh(E^{z})$ is the revealed information in classical error correction.
The gain $Q^{z}$ and quantum bit error rate $E^{z}$ of the $Z$ basis can be directly acquired in experiment, where $E^{z}$ is the maximum value between the marginal bit error rates of Alice-Charlie and Bob-Charlie when the raw key of Charlie is set as the reference key. The yields $Y_{0}^{d}=Q_{00}^{d}$ and $Y_{1}^{d}$ can be estimated by exploiting the decoy-state method~\cite{wang2005beating,lo2005decoy},
\begin{equation}
\begin{aligned}
Y_{1}^{d}\geq \frac{\mu/2}{\mu\nu-\nu^{2}}\left(e^{\nu}Q_{\nu}^{d}-\frac{\nu^{2}}{\mu^{2}}e^{\mu}Q_{\mu}^{d}-\frac{\mu^{2}-\nu^{2}}{\mu^{2}}Q_{0}^{d}\right),
\end{aligned}
\end{equation}
where we define $Q_{k}^{d}=Q_{k0}^{d}+Q_{0k}^{d}$. $Q_{k_{a}k_{b}}^{d}$ is the gain of Charlie selecting the $d$ basis given that Alice and Bob send intensities $k_{a}$ and $k_{b}$ and can be given by
\begin{equation}
\begin{aligned}
Q_{k_{a}k_{b}}^{z}=&(1-p_{d})(e^{-k_{a}\sqrt{\eta}}+e^{-k_{b}\sqrt{\eta}})\\
&-2(1-p_{d})^{2}e^{-(k_{a}+k_{b})\sqrt{\eta}}.
\end{aligned}
\end{equation}
Besides, the gain $Q_{k_{a}k_{b}}^{x}$ can be written as
\begin{equation}
\begin{aligned}
Q_{k_{a}k_{b}}^{x}=&2(1-p_{d})e^{-\frac{k_{a}+k_{b}}{2}\sqrt{\eta}}
\times[I_{0}(\sqrt{k_{a}k_{b}}\sqrt{\eta})\\
&-(1-p_{d})e^{-\frac{k_{a}+k_{b}}{2}\sqrt{\eta}}],
\end{aligned}
\end{equation}
where $I_{0}(x)$ is the modified Bessel function of the first kind.

For $Y_{1}^{x}$,  we employ the fact that $\frac{1}{2}(\ket{10}_{ab}\bra{10}+\ket{01}_{ab}\bra{01})=\frac{1}{2}(\ket{\psi^{+}}_{ab}\bra{\psi^{+}}_{ab}+\ket{\psi^{-}}_{ab}\bra{\psi^{-}}_{ab})$.
The bit error rate $e_{1}^{x}$ can be bounded by

\begin{equation}
\begin{aligned}
e_{1}^{x}\leq \frac{1}{2\nu Y_{1}^{x}p_{\rm pm}}\left(e^{2\nu}\mathsf{E}_{\nu\nu}^{x}\mathsf{Q}_{\nu\nu}^{x}-\frac{p_{\rm pm}}{2}Y_{0}^{x}\right),
\end{aligned}
\end{equation}
where $\mathsf{E}_{\nu\nu}^{x}$ and $\mathsf{Q}_{\nu\nu}^{x}$ are the bit error rate and gain of Charlie selecting the $X$ basis when Alice and Bob both send intensity $\nu$ and they successfully perform the post-selected phase-matching.

Note that quantum bit error rate $E^{z}$ is the maximum value between the marginal error rates of Alice-Charlie and Bob-Charlie as Charlie's raw key is the reference key in the error correction step. For the case of our practical protocol, we have the gain $Q^{z}=Q_{c}^{z}+Q_{e}^{z}$ and quantum bit error rate $E^{z}=Q_{e}^{z}/Q^{z}$. Therein, $Q_{c}^{z}$ and $Q_{e}^{z}$ are the correct and incorrect gains between Alice and Bob,
\begin{equation}
\begin{aligned}\label{}
Q_{c}^{z}&=t(1-t)(Q_{\mu0}^{z}+Q_{0\mu}^{z}),\\
Q_{e}^{z}&=(1-t)^{2}Q_{00}^{z}+t^{2}Q_{\mu\mu}^{z},
\end{aligned}
\end{equation}

For the case of Alice and Bob both choosing intensity $\nu$, they should perform the post-selected phase-matching.
The corresponding gain and error rate can be written as $\mathsf{Q}_{\nu\nu}^{x}=\mathsf{Q}_{c,\nu\nu}^{x}+\mathsf{Q}_{e,\nu\nu}^{x}$ and $\mathsf{E}_{\nu\nu}^{x}=[e_{d}^{x}\mathsf{Q}_{c,\nu\nu}^{x}+(1-e_{d}^{x})\mathsf{Q}_{e,\nu\nu}^{x}]/\mathsf{Q}_{\nu\nu}^{x}$. Therein, the correct gain $\mathsf{Q}_{c,\nu\nu}^{x}$ and incorrect gain $\mathsf{Q}_{e,\nu\nu}^{x}$ are
\begin{equation}
\begin{aligned}
\mathsf{Q}_{c,\nu\nu}^{x}=&\frac{p_{\rm pm}(1-p_{d})}{\delta}\int_{0}^{\delta}e^{-\nu\sqrt{\eta}(1-\cos \theta)}d\theta\\
&-p_{\rm pm}(1-p_{d})^{2}e^{-2\nu\sqrt{\eta}}\\
\approx&\frac{p_{\rm pm}(1-p_{d})}{\delta}\sqrt{\frac{\pi}{2\nu\sqrt{\eta}}}{\rm erf}\left(\sqrt{\frac{\nu\sqrt{\eta}}{2}}\delta\right)\\
&-p_{\rm pm}(1-p_{d})^{2}e^{-2\nu\sqrt{\eta}},\\
\end{aligned}
\end{equation}
and
\begin{equation}
\begin{aligned}\label{eq6}
\mathsf{Q}_{e,\nu\nu}^{x}=&\frac{p_{\rm pm}(1-p_{d})}{\delta}\int_{0}^{\delta}e^{-\nu\sqrt{\eta}(1+\cos \theta)}d\theta\\
&-p_{\rm pm}(1-p_{d})^{2}e^{-2\nu\sqrt{\eta}}\\
\approx&\frac{p_{\rm pm}(1-p_{d})}{\delta}e^{-2\nu\sqrt{\eta}}\sqrt{\frac{\pi}{2\nu\sqrt{\eta}}}{\rm erfi}\left(\sqrt{\frac{\nu\sqrt{\eta}}{2}}\delta\right)\\
&-p_{\rm pm}(1-p_{d})^{2}e^{-2\nu\sqrt{\eta}},\\
\end{aligned}
\end{equation}
where the approximate results are always true for $0<\delta\leq\frac{\pi}{4}$. Therein, ${\rm erf}(x)$ and ${\rm erfi}(x)$ are the error function and imaginary error function.

\noindent
\textbf{Single-photon protocol.}\label{ent}
Here, we also present an ideal protocol, called the single-photon protocol, where ideal single-photon sources are required to replace the phase-randomized weak coherent pulses in the $Z$ basis. Alice (Bob) randomly sends optical modes $\ket{0}$ and $\ket{1}$ by using the vacuum and single-photon source with probabilities $1-t$ and $t$ in the $Z$ basis. 

Following the entanglement distillation argument~\cite{fu2015long}, the asymptotic conference key rate of the single-photon protocol is given by
\begin{equation}
\begin{aligned}
\widetilde{R}=2t(1-t)Y_{1}^{z}[1-h(e_{1}^{x})]-\widetilde{\lambda}_{\rm EC},
\end{aligned}
\end{equation}
where $Y_{1}^{d}$ and $e_{1}^{d}$ are the yield and bit error rate of the joint single-photon state given that all users choose the $d$ basis, with $d\in\{Z,X\}$. For example, $Y_{1}^{z}$ is the probability that Charlie has a successful detection in the $Z$ basis given that Alice and Bob send the state  $\rho_{ab}^1$. Let $x_{a}\oplus x_{b}\neq x_{c}$ denote the bit error of the $X$ basis, where $x_{c}$ is the logic bit value of Charlie in the $X$ basis.
$\widetilde{\lambda}_{\rm EC}$ is the leaked information in classical error correction. $h(x)=-x\log_{2}(x)-(1-x)\log_{2}(1-x)$ is the binary Shannon entropy. There is always a marginal bit error rate in the $Z$ basis between every two users. Taking the raw key of Charlie as the reference key, the $Z$ basis bit error rate of this protocol is the maximum value between the marginal bit error rates of Alice-Charlie and Bob-Charlie.

For the case of the scheme with single-photon source, Alice and Bob directly prepare optical mode $\ket{1}$ with single photon.
We also have the gain $Q^{z}=Q_{c}^{z}+Q_{e}^{z}$ and quantum bit error rate $E^{z}=Q_{e}^{z}/Q^{z}$. Therein, $Q_{c}^{z}$ and $Q_{e}^{z}$ are the correct and incorrect gains between Alice and Bob,
\begin{equation}
\begin{aligned}
Q_{c}^{z}&=t(1-t)(Y_{10}^{z}+Y_{01}^{z}),\\
Q_{e}^{z}&=(1-t)^{2}Y_{00}^{z}+t^{2}Y_{11}^{z},
\end{aligned}
\end{equation}
where these yields can be expressed as
\begin{equation}
\begin{aligned}\label{}
Y_{10}^{z}&=Y_{01}^{z}=1-(1-p_{d})(1-2p_{d})(1-\sqrt{\eta}),\\
Y_{11}^{z}&=2(1-p_{d})(1-\sqrt{\eta})[1-(1-p_{d})(1-\sqrt{\eta})],
\end{aligned}
\end{equation}
and $Y_{00}^{z}=2p_{d}(1-P_{d})$. Thereby, the revealed information in classical error correction is $\widetilde{\lambda}_{\rm EC}=Q^{z}fh(E^{z})$. Besides, we have the yields $Y_{1}^{z}=\frac{1}{2}(Y_{10}^{z}+Y_{01}^{z})$. Note that $Y_{1}^{x}=\frac{1}{2}(Y_{10}^{x}+Y_{01}^{x})$ due to $\frac{1}{2}(\ket{10}_{ab}\bra{10}+\ket{01}_{ab}\bra{01})=\frac{1}{2}(\ket{\psi^{+}}_{ab}\bra{\psi^{+}}_{ab}+\ket{\psi^{-}}_{ab}\bra{\psi^{-}}_{ab})$, where $Y_{10}^{x}$ and $Y_{01}^{x}$ can be given by
\begin{equation}
\begin{aligned}
Y_{10}^{x}=Y_{01}^{x}=1-(1-p_{d})(1-2p_{d})(1-\sqrt{\eta}).
\end{aligned}
\end{equation}

\begin{figure}
\centering
\includegraphics[width=8cm]{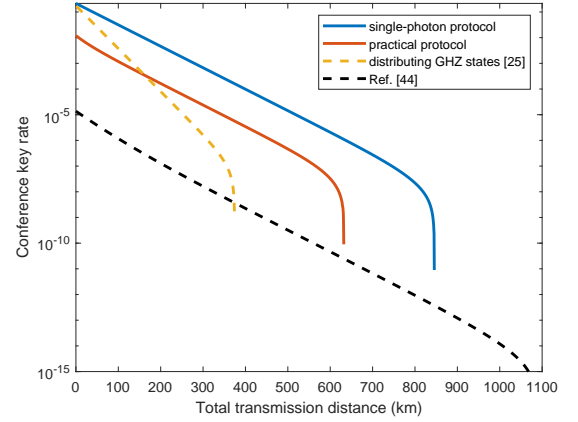}
    \caption{Conference key rate (per pulse) in logarithmic scale as a function of the total transmission distance using the experimental parameters in Table~\ref{tab1}. For comparison, we also draw the curves of the single-photon protocol, Ref.~\cite{Zhao:2020:CKA} and distributing GHZ states among three parties~\cite{epping2017multi}, where the transmitting distance to Charlie is zero while the transmitting distances to Alice and Bob are $L/2$.
} \label{f2}
\end{figure}

\section{Performance}\label{Sec_Dis}
We assume that the efficiencies and dark count rates of Charlie's detectors are the same and the distances of Alice-Charlie and Bob-Charlie are both $L/2$. We numerically optimize the conference key rate over the free parameters $t$, $\mu$ and $\nu$ by utilizing Genetic Algorithm. To display the conference key rate of our protocol scaling with the square-root of the total channel transmittance, we define $\sqrt{\eta}=\eta_{d}\times10^{-\alpha L/20}$, where $\eta_{d}$ is the detector efficiency. In order to show the performance of our quantum CKA protocol, we utilize the practical parameters in Table~\ref{tab1} for simulation, which has been recently realized over 500 km fiber in twin-field QKD~\cite{chen2020sending}. The conference key rate of our practical quantum CKA protocol as a function of the total transmission distance $L$ among three users is shown in Fig.~\ref{f2}. The transmission distance of our practical protocol reaches over 600 km and its key rate can surpass that of distributing GHZ states among 3 parties~\cite{epping2017multi} at the distance of 160 km. For the case of distributing GHZ states~\cite{epping2017multi}, we set the misalignment rate of its channels is 0 and only dark counts contribute to its error rate, which aims to make its key rate close to the GHZ entanglement distribution capacity. Compared with the scheme of Ref.~\cite{Zhao:2020:CKA}, although our scheme is not a measurement-device-independent protocol, unlike Ref.~\cite{Zhao:2020:CKA}, the key rate of our protocol has improved significantly and is three orders of magnitude higher within 600 km. Additionally, the single-photon protocol can be theoretically demonstrated over 800 km, which also shows the importance to develop ideal single-photon source.

\begin{figure}
\centering
\includegraphics[width=8cm]
{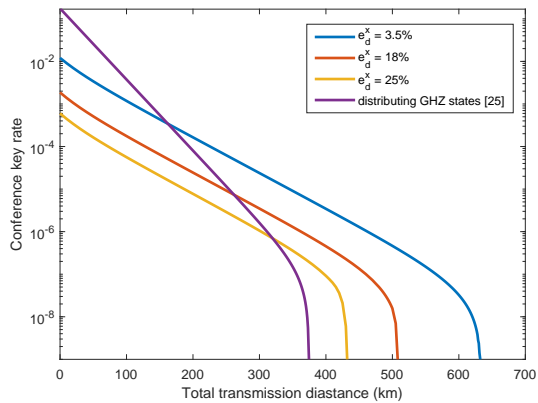}
\caption{Conference key rates of the practical protocol in logarithmic scale as a function of the total distance with different misalignment rates of the $X$ basis.
The conference key rate can also beat that of distributing GHZ states among 3 parties~\cite{epping2017multi}  even when the misalignment rate is up to $e_{d}^{x}=25\%$.
} \label{f3}
\end{figure}

Besides, the bit error rate $E^{z}$ can be bounded by adjusting probability $t$, which is irrelevant to the misalignment of channel. Therefore, our practical protocol can always extract conference key as long as the misalignment rate of $X$ basis $e_{d}^{x}<50\%$ in principle. As shown in Fig.~\ref{f3}, the secure transmission distance is larger than 500 km in the case of $e_{d}^{x}=18\%$. Besides, the conference key rate of the practical protocol can surpass that of distributing GHZ states over the networks~\cite{epping2017multi} even when the misalignment rate is larger than $25\%$. These results indicate that our quantum CKA is practical and feasible even in the field environment.

\begin{table}
\centering
\caption{Simulation parameters~\cite{chen2020sending}. $\eta_{d}$ and $p_{d}$ are the detector efficiency and dark count rate. $e_{d}^{x}$ is the misalignment rate of the $X$ basis. $\alpha$ is the attenuation coefficient of the ultralow-loss fiber. $f$ is the error correction efficiency.}\label{tab1}
\begin{tabular}{cccccc}
\hline
\hline
$\eta_{d}$ ~~~~~& $p_{d}$ ~~~~~& $e_{d}^{x}$~~~~~ & $\alpha$ ~~~~~& $f$&~~~~~ $\delta$\\
\hline
$56\%$~~~~~& $10^{-8}$~~~~~ & $3.5\%$ ~~~~~ & $0.167$ ~~~~~& $1.1$&~~~~~$\pi/18$\\
\hline
\hline
\end{tabular}
\end{table}

\section{Conclusion}
In summary, by employing a special GHZ-class state, we have proposed a practical quantum CKA protocol that allows three users to share the information-theoretically secure conference key.
Its conference key rate scales as $O(\sqrt{\eta})$ rather than the total transmittance, which can beat the fundamental linear limit on the private capacity of quantum network~\cite{das2019universal}.
This protocol exploits the same devices and technology as twin-field QKD~\cite{lucamarini2018overcoming,minder2019experimental,chen2020sending,fang2020implementation} and as such, it can be demonstrated over 600 km. New results are significantly beyond what one could expect for quantum CKA in previous works, whose key rate is three orders of magnitude higher within 600 km compared to Ref.~\cite{Zhao:2020:CKA} and keeps at a high level even with large misalignment rate. We believe that this quantum CKA protocol can be widely implemented to build the large-scale quantum cryptographic network. 

Here, we would like to clarify that our quantum CKA is not a measurement-device-independent scheme and will suffer from the standard quantum hacking attacks, including blinding attack~\cite{Lydersen:2010:Hacking}. In order to treat double-click attack, one should exploit the squashing model~\cite{gottesman2004security}, i.e., random basis and random measurement outcomes should be assigned. Our protocol also requires the phase stability of two independent pulses, which can be circumvented by using active phase stabilisation~\cite{Wang:2019:Beating, minder2019experimental}, phase tracking and post-selection~\cite{Liu:2019:Exp}. Different from the scheme with post-selected W state~\cite{grasselli2019conference}, our scheme with arbitrary $N$ parties is an interesting and nontrivial work.

\section*{Acknowledgments}
We gratefully acknowledge support from the National Natural Science Foundation of China (under Grant No. 61801420); the Key-Area Research and Development Program of Guangdong Province (under Grant No. 2020B0303040001); the Fundamental Research Funds for the Central Universities (under Grant No. 020414380182).

\section*{Disclosures}
\noindent The authors declare no conflicts of interest.


\end{document}